\documentclass[aps,prl,twocolumn,groupedaddress,nofootinbib,showpacs,preprintnumbers]{revtex4}%

\usepackage{graphicx}

\usepackage{amsmath}
\usepackage{slashed}
\usepackage{amssymb}
\usepackage{bm}
\usepackage{color}

\usepackage[hypertex]{hyperref}

\begin{document}


\preprint{DESY 18-168\hspace{13.8cm}ISSN 0418--9833}
\preprint{September 2018\hspace{15.75cm}}


\title{Breakdown of Nonrelativistic QCD Factorization in Processes Involving
  Two Quarkonia and its Cure}

\author{Zhi-Guo He, Bernd A. Kniehl, Xiang-Peng Wang}

\affiliation{ {II.} Institut f\"ur Theoretische Physik, Universit\"at Hamburg,
Luruper Chaussee 149, 22761 Hamburg, Germany}

\date{\today}

\begin{abstract}
We study inclusive processes involving two heavy quarkonia in nonrelativisitic
QCD (NRQCD) and demonstrate that, in the presence of two P-wave Fock states,
NRQCD factorization breaks down, leaving uncanceled infrared singularities.
As phenomenologically important examples, we consider the decay
$\Upsilon \to \chi_{cJ}+X$ via
$b\bar{b}({}^3P_{J_b}^{[8]})\to c\bar{c}({}^3P_J^{[1]})+gg$
and the production process
$e^+e^-\to J/\psi+\chi_{cJ}+X$ via
$e^{+}e^{-}\to c\bar{c}({}^3P_{J_1}^{[8]})+c\bar{c}({}^3P_J^{[1]})+g$.
We infer that such singularities will appear for double quarkonium
hadroproduction at next-to-leading order.
As a solution to this problem, we introduce to NRQCD effective field theory new
types of operators whose quantum corrections absorb these singularities.
\end{abstract}




\maketitle


Because of their large mass scales and nonrelativistic nature, heavy-quarkonium
states are ideal probes to study quantum chromodynamics (QCD), which is the
fundamental theory to describe the strong interactions between quarks and
gluons, from both perturbative and nonperturbative aspects.
The traditional color-singlet model (CSM), which restricts the heavy-quark
pair, $Q\bar{Q}$, to form a color singlet and to share the spectroscopic
quantum numbers ${}^{2S+1}L_J$ of spin $S$, orbital angular momentum $L$, and
total angular momentum $J$ with the physical quarkonium state, is plagued by
infrared (IR) singularities when applied to the production or decay of P-wave
quarkonia \cite{Barbieri:1976fp} or quarkonia with $L>1$
\cite{Belanger:1987cg}.
Phenomenologically, a cutoff, e.g., on the binding energy of the $Q\bar{Q}$
bound state or the momenta of emitted gluons, has to be introduced to
regularize such singularities, which makes the theoretical predictions
model dependent and causes the separation of short- and long-distance physics to
break down.
This problem has been successfully solved by the factorization formalism
\cite{Bodwin:1994jh} built on the rigorous effective field theory of
nonrelativistic QCD (NRQCD) \cite{Caswell:1985ui}.
In the NRQCD factorization formalism \cite{Bodwin:1994jh}, the production and
decay rates of heavy quarkonia are separated into short-distance coefficients
(SDCs), which can be obtained as expansions in the strong-coupling constant
$\alpha_s$ through the NRQCD to full-QCD matching of perturbative calculations,
and supposedly universal long-distance matrix elements (LDMEs).
The sizes of the latter are subject to scaling rules in the velocity $v_Q$
of $Q$ and $\bar{Q}$ in the $Q\bar{Q}$ rest frame \cite{Lepage:1992tx}.
This allows us to calculate heavy-quarkonium production and decay rates 
systematically as double expansions in $\alpha_s$ and $v_Q^2$.
The NRQCD factorization formalism has successfully cured the IR problem of the
CSM, as explicitly shown in the literature for inclusive decay and production
of P- \cite{Huang:1996fa} and D-wave states \cite{Fan:2009cj}.
Note that in exclusive processes, such as $B$-meson exclusive decay to
$\chi_{cJ}$ mesons \cite{Song:2003yc} and exclusive double quarkonium
production in $e^{+}e^{-}$ annihilation \cite{Bodwin:2008nf}, there are still
uncanceled IR divergences at one loop, which disappear in the limits
$m_c/m_b\to0$ and $m_c/\sqrt{s}\to0$, respectively \cite{Bodwin:2008nf}.

Besides inclusive production and decay processes involving a single quarkonium
state, also processes in which two heavy quarkonia participate, such as
bottomonium decay to charmonium or double heavy-quarkonium production, are of
great phenomenological interest.
Measurements of $\Upsilon$ decay to charmonium, like the $J/\psi$ meson, can be
traced to the first experiment carried out by CLEO Collaboration about 30 years
ago \cite{Fulton:1988ug}, and were then updated by the ARGUS 
\cite{Albrecht:1992ap} and CLEO \cite{Briere:2004ug} Collaborations with much
larger data samples. 
Very recently, more precise results for the branching ratios of
$\Upsilon\to J/\psi(\psi^{\prime})+X$ \cite{Shen:2016yzg} and
$\Upsilon\to\chi_{c1}+X$ \cite{Jia:2016cgl} were obatined by the Belle
Collaboration.
On the other hand, $J/\psi$ pair and $J/\psi+\Upsilon$ associated production
have been a very hot topic at hadron colliders in recent years.
In fact, double $J/\psi$ ($\Upsilon$) and $J/\psi+\Upsilon$ prompt production
were extensively measured by the D0 Collaboration \cite{Abazov:2014qba} at the
FNAL Tevatron and by the LHCb \cite{Aaij:2011yc}, CMS
\cite{Khachatryan:2014iia}, and ATLAS \cite{Aaboud:2016fzt} Collaborations at
the CERN LHC.
Interestingly, there are substantial discrepancies between CMS data and NRQCD
predictions at leading order (LO) in $\alpha_s$, which are expected to be
reduced by the yet unknown next-to-leading-order (NLO) corrections
\cite{He:2015qya}.
Moreover, double heavy-quarkonium production serves as a useful laboratory to
investigate the double parton scattering mechanism \cite{Kom:2011bd} at hadron
colliders.
On the theoretical side, both bottomonium decay to charmonium and double
heavy-quarkonium production have been studied in the NRQCD factorization
framework. 
In the former case, only the color-singlet (CS) ${}^3S_1^{[1]}$
\cite{Trottier:1993ze} and color-octet (CO) ${}^3S_1^{[8]}$
\cite{Cheung:1996mh} $b\bar{b}$ Fock states were considered.
In the latter case, the CS contributions are known to NLO in $\alpha_s$
\cite{Sun:2014gca} and $v_Q^2$ \cite{Li:2013csa}, while the CO
contributions are only known to LO \cite{He:2015qya} and partially to NLO in
$v_Q^2$ \cite{Li:2013csa}.
In all these calculations, IR singularities appearing in intermediate steps
were always properly removed by NRQCD factorization, and one might have
expected that this is a general rule valid for all $Q\bar{Q}$ Fock states and
to all orders in  $\alpha_s$ and $v_Q^2$.

In this Letter, we change this familiar picture by presenting two
counterexamples, suggesting that the well-established formalism needs a
generalization for the cases at hand.
In fact, the violation of NRQCD factorization appears as one includes more
$Q\bar{Q}$ Fock states or goes beyond LO in $\alpha_s$.
As we demonstrate later on, the left-over IR singularities may, fortunately, be
completely absorbed into the QCD corrections to a class of operators introduced
here for the first time. 

Let us first consider the inclusive production of charmonium
$H=J/\psi,\chi_{cJ},\psi^{\prime}$ by $\Upsilon$ decay.
By NRQCD factorization, the decay width can be expressed as 
\begin{eqnarray}
\lefteqn{\Gamma(\Upsilon\to H+X)}\nonumber\\
&&{}=\sum_{m,n}\hat{\Gamma}(b\bar{b}(m)\to c\bar{c}(n)+X)
\langle\Upsilon|\mathcal{O}(m)|\Upsilon\rangle
\langle\mathcal{O}^{H}(n)\rangle,\qquad
\end{eqnarray}
where $m$ and $n$ are $b\bar{b}$ and $c\bar{c}$ Fock states, respectively,
$\hat{\Gamma}(b\bar{b}(m)\to c\bar{c}(n)+X)$ are the SDCs, and
$\langle\Upsilon|\mathcal{O}(m)|\Upsilon\rangle$ and
$\langle\mathcal{O}^{H}(n)\rangle$ are the LDMEs. 
According to the velocity scaling rules \cite{Lepage:1992tx}, for $\Upsilon$
($J/\psi$, $\psi^{\prime}$) the four Fock states ${}^3S_1^{[1]}$,
${}^1S_0^{[8]}$, ${}^3S_1^{[8]}$, and ${}^3P_J^{[8]}$ contribute through
relative order $v_b^4$ ($v_c^4$), while for $\chi_{cJ}$ the dominant
contributions come from ${}^3P_J^{[1]}$ and ${}^3S_1^{[8]}$ at order $v_c^2$.
We have checked explicitly that, in the single P-wave case, where either
$b\bar{b}$ or $c\bar{c}$ is in a P-wave state, the IR singularities arising in
the full QCD calculation can be completely absorbed into the corresponding
S-wave LDMEs rendering $\hat{\Gamma}(b\bar{b}(m)\to c\bar{c}(n)+X)$ IR finite.
However, when both $b\bar{b}$ and $c\bar{c}$ are in P-wave states, there are
extra IR singularities that cannot be absorbed into the S-wave LDMEs.
At LO, there are two such subprocesses, namely
$b\bar{b}({}^3P_{J_b}^{[8]})\to c\bar{c}({}^3P_{J_c}^{[8]})+gg$ for
$\Upsilon\to J/\psi+X$ and 
$b\bar{b}({}^3P_{J_b}^{[8]})\to c\bar{c}({}^3P_{J_c}^{[1]})+gg$ for
$\Upsilon\to\chi_{cJ}+X$.
In the following, we focus our attention on the second one.
Similar conclusions can be drawn for the first one and eventually be extended
to the general case of bottomonium decay to charmonium plus anything.

\begin{figure}
\begin{center}
\includegraphics[width=0.48\textwidth]{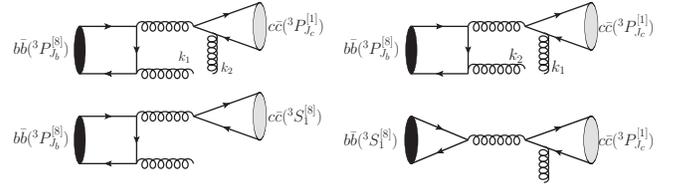}
\caption{Typical Feynman diagrams for
$b\bar{b}({}^3P_{J_b}^{[8]})\to c\bar{c}({}^3P_{J_c}^{[1]})+gg$ (upper panel),
$b\bar{b}({}^3P_{J_b}^{[8]})\to c\bar{c}({}^3S_1^{[8]})+g$ (lower left panel),
and $b\bar{b}({}^3S_1^{[8]})\to c\bar{c}({}^3P_{J_c}^{[1]})+g$ (lower right
panel).}
\label{Upsilon_Decay} 
\end{center}
\end{figure}

There are eight Feynman diagrams for
$b\bar{b}({}^3P_{J_b}^{[8]})\to c\bar{c}({}^3P_{J_c}^{[1]})+gg$, and typical
ones are depicted in Fig.~\ref{Upsilon_Decay}.
For each choice of $J_b$ and $J_c$, the corresponding partonic decay rate may
be calculated by using appropriate spin and color projectors.
Unsurprisingly, they all contain IR-divergent terms. 
Because of space limitation, we present here only the latter.
Furthermore, we sum over $J_b$ using heavy-quark spin symmetry, leaving 
$\langle\Upsilon|\mathcal{O}({}^3P_{0}^{[8]})|\Upsilon\rangle$ as the overall
$\Upsilon$ LMDE.
We write the result as
\begin{equation}
  \hat{\Gamma}^{\mathrm{div}}(J_c)=\hat{\Gamma}_1^{\mathrm{div}}
  +9\hat{\Gamma}_2^{\mathrm{div}}(J_c)+\hat{\Gamma}_3^{\mathrm{div}}(J_c),
\end{equation}
where $\hat{\Gamma}_1^{\mathrm{div}}$ ($\hat{\Gamma}_2^{\mathrm{div}}(J_c)$)
arises from the square of the amplitude ${\cal M}_1$ (${\cal M}_2$) of the
diagrams in which the soft gluon is emitted by the $c$ or $\bar{c}$
($b$ or $\bar{b}$) quarks, and $\hat{\Gamma}_3^{\mathrm{div}}(J_c)$ arises from
the interference of ${\cal M}_1$ and ${\cal M}_2$.
We have:
\begin{eqnarray}
\hat{\Gamma}_1^{\mathrm{div}}&=&\frac{-8\alpha_s}{27\pi m_c^2}\frac{1}{\epsilon_{\mathrm{IR}}}\times
\frac{5\pi^2\alpha_s^3(3r^4+2r^2+7)}{72m_b^7r^3(1-r^2)},
\nonumber\\
\hat{\Gamma}_2^{\mathrm{div}}(J_c)&=&\frac{-5\alpha_s}{9\pi m_b^2}
\frac{1}{\epsilon_{\mathrm{IR}}}\times
\begin{cases} 
\frac{\pi ^2 \alpha_s^3 \left(1-3 r^2\right)^2}{81 m_b^7 r^3 
\left(1-r^2\right)}, & J_c=0,\\
\frac{2 \pi ^2 \alpha_s^3 \left(r^2+1\right)}{81 m_b^7 r^3 \left(1-r^2\right)},
 & J_c=1,\\
\frac{2 \pi ^2 \alpha_s^3 \left(6 r^4+3 r^2+1\right)}{405 m_b^7 r^3 \left(1-r^2\right)}, & J_c=2,
\end{cases}
\nonumber\\
\hat{\Gamma}_{3}^{\text{div}}(0)&=&-\frac{10 \pi \alpha_s^4}{81 m_b^9 r^3 \left(1-r^2\right)^4 \epsilon_{\mathrm{IR}} }\nonumber\\
&&{}\times\left(3r^4-10 r^2+3\right) \left(r^4-4 r^2 \ln r-1\right),
\nonumber\\
\hat{\Gamma}_{3}^{\text{div}}(1)&=&\frac{10\pi \alpha_s^4}{81 m_b^9 r^3 \left(1-r^2\right)^4 \epsilon_{\mathrm{IR}}  }\left[-r^6+9r^4\right. \nonumber\\
&&{}\left.-7 r^2+4 r^2 \left(r^4-3 r^2-2\right) \ln r-1\right],
\nonumber\\
\hat{\Gamma}_{3}^{\text{div}}(2)&=&\frac{2 \pi  \alpha_s^4}{81 m_b^9 r^3 \left(1-r^2\right)^4 \epsilon_{\mathrm{IR}}  }\left[6 r^8+23 r^6\right.\nonumber\\
&&{}\left.-27 r^4+r^2-4 r^4\left(9 r^2+11 \right)  \ln r-3\right],
\label{eq:gamma}
\end{eqnarray}
where $\epsilon_{\mathrm{IR}}=D/2-2$ is the infrared regulator of dimensional
regularization and $r=m_c/m_b$.
In Eq.~(\ref{eq:gamma}), we write $\hat{\Gamma}_1^{\mathrm{div}}$ and
$\hat{\Gamma}_2^{\mathrm{div}}(J_c)$ as products of the SDCs of
$b\bar{b}({}^3P_{J_b}^{[8]})\to c\bar{c}({}^3S_1^{[8]})+g$ and
$b\bar{b}({}^3S_{1}^{[8]})\to c\bar{c}({}^3P_{J_c}^{[1]})+g$,  
whose representative Feynman diagrams are shown in Fig.~\ref{Upsilon_Decay} as
well, and the IR-singular terms arising from the NLO QCD corrections to the
LDMEs $\langle\mathcal{O}^{\chi_{cJ}}({}^3S_{1}^{[8]})\rangle$ and 
$\langle \Upsilon|\mathcal{O}({}^3S_{1}^{[8]})|\Upsilon\rangle$, respectively,
so as to indicate that they will be canceled after taking into account the
contributions of $b\bar{b}({}^3P_{J_b}^{[8]})\to c\bar{c}({}^3S_1^{[8]})+g$ and
$b\bar{b}({}^3S_{1}^{[8]})\to c\bar{c}({}^3P_{J_c}^{[1]})+g$.
However, in the case of $\hat{\Gamma}_3^{\mathrm{div}}(J_c)$, the NRQCD
factorization formalism as we know it simply lacks an operator that could
compensate the soft-gluon effects.
This renders $\hat{\Gamma}^{\mathrm{div}}(J_c)$ IR singular altogether.
In other words, we are faced by an IR problem of NRQCD factorization which has
gone unnoticed so far!
Since $\hat{\Gamma}_3^{\mathrm{div}}(J_c)$ are due to the interference of
diagrams with soft gluons emitted by P-wave $b\bar{b}$ and $c\bar{c}$ Fock
states, which appears in NRQCD treatments of any inclusive decay of bottomonium
to charmonium at some order of $v_b^2$ and $v_c^2$, we conclude that the NRQCD 
factorization formalism, in its familiar and generally accepted form, will
break down for any such process.
In some cases, this may happen even at relative order $v_{b,c}^2$, i.e.\ at
LO, e.g.\ for the decay $\chi_{b{J_b}}\to\chi_{c{J_c}}+X$ via the channel
$b\bar{b}({}^3P_{J_b}^{[1]})\to c\bar{c}({}^3P_{J_c}^{[1]})+gg$.
We note in passing that, unlike for the exclusive processes mentioned above
\cite{Song:2003yc,Bodwin:2008nf}, $\hat{\Gamma}_{3}^{\text{div}}(J_c)$ does not
vanish in the limit $r\to0$.

\begin{figure}
\begin{center}
\includegraphics[width=0.48\textwidth]{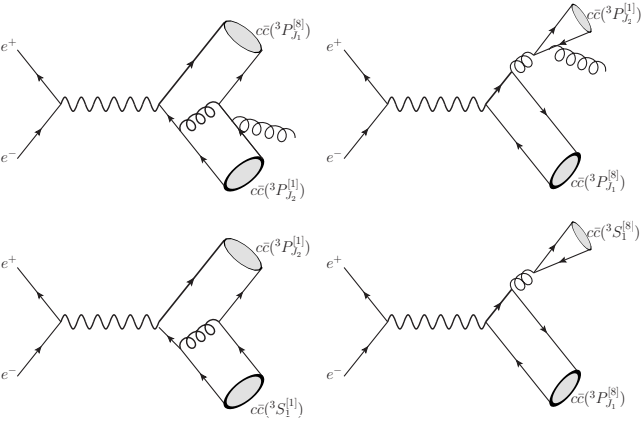}
\caption{Typical Feynman diagrams for
  $e^{+}e^{-}\to c\bar{c}({}^3P_{J_1}^{[8]})+c\bar{c}({}^3P_{J_2}^{[1]})+g$
  (upper panel),
  $e^{+}e^{-}\to c\bar{c}({}^3P_{J_1}^{[8]})+c\bar{c}({}^3S_1^{[8]})$
  (lower left panel), and
  $e^{+}e^{-}\to c\bar{c}({}^3S_{1}^{[1]})+c\bar{c}({}^3P_{J_2}^{[1]})$
  (lower right panel).}
\label{epem} 
\end{center}
\end{figure}

We now turn to our second example, the inclusive production of two heavy
quarkonia, e.g.\ $2J/\psi$, $J/\psi+\Upsilon$, etc.
In the NRQCD treatment of $J/\psi$ pair or $J/\psi+\Upsilon$ associated
hadroproduction, soft-gluon emission starts from NLO in $\alpha_s$, e.g.\
$gg\to c\bar{c}({}^3P_{J_1}^{[8]})+c\bar{c}({}^3P_{J_2}^{[8]})+g$. 
A complete NLO NRQCD calculation lies beyond the scope of this Letter.
Instead, we choose a relatively simple process for illustration,
namely $e^{+}e^{-}\to J/\psi+\chi_{cJ}+X$ proceeding via the LO channel
$e^{+}e^{-}\to c\bar{c}({}^3P_{J_1}^{[8]})+c\bar{c}({}^3P_{J_2}^{[1]}) + g$.
The IR problem featured here should also appear in double $J/\psi$ production
via $e^{+}e^{-}$ annihilation \cite{Feng:2017bdu}.

There are 28 contributing Feynman diagrams, typical ones of which are
displayed in Fig.~\ref{epem}.
Again, there are three sources of IR singularities in the full NRQCD
calculation:
(1) the square of the part of the amplitude where the gluon is attached to the
$c\bar{c}({}^3P_{J_1}^{[8]})$ pair, (2) the same for ${}^3P_{J_2}^{[1]}$,
and (3) the interference of these two amplitude parts.
After summing over $J_1$, we have for the IR-singular pieces
\begin{equation}
\hat{\sigma}^{\mathrm{div}}(J_2)=\sigma^{\mathrm{div}}_{1}
+9\sigma^{\mathrm{div}}_{2}(J_2)+\sigma^{\mathrm{div}}_{3}(J_2),
\end{equation}
with
\begin{widetext}
\begin{eqnarray}
  \hat{\sigma}_1^{\mathrm{div}} &=& -\frac{8\alpha_s}{27\pi m_c^2}
  \frac{1}{\epsilon_{\mathrm{IR}}}\times
\frac{2^{10}\pi^3\alpha^2\alpha_s^2S}{3^6 s^5 r^6}
(864r^{10}-144r^8-1568r^6+1224r^4-130r^2+27),
\nonumber\\
\hat{\sigma}^{\mathrm{div}}_2(J_2) &=&-\frac{4\alpha_s}{3\pi m_c^2}
\frac{1}{\epsilon_{\mathrm{IR}}}\times
\frac{2^{18}\pi^3\alpha^2\alpha_s^2S}{3^9 s^5 r^4}
\begin{cases} 
(144r^8+152 r^6-428r^4+182r^2+1), & J_2=0, \\
8(18r^6+13r^4-12r^2+2), & J_2=1,\\
\frac{2}{5}(360r^8+308r^6-188r^4+20r^2+1), & J_2=2,
\end{cases}
\nonumber\\
\hat{\sigma}^{\text{div}}_{3}(0)&=&
\frac{2^{18} \pi^2\alpha^2\alpha_s^3}{3^8 s^6 r^4\epsilon_{\mathrm{IR}}}
\left[\left(144 r^8+184 r^6-504 r^4+170 r^2+33\right)S
  +8 \left(72 r^{10}+56 r^8-284 r^6+149 r^4+r^2\right)T\right],
\\
\hat{\sigma}^{\text{div}}_{3}(1)&=&
\frac{2^{18} \pi^2\alpha^2\alpha_s^3}{3^8 s^6 r^2\epsilon_{\mathrm{IR}}} 
\left[\left( 144 r^6+28 r^4-176r^2+43\right)S
  +\left(576 r^{10}-176 r^8-792 r^6+424 r^4-48r^2\right)T\right],
\nonumber\\
\hat{\sigma}^{\text{div}}_{3}(2)&=&
\frac{2^{18} \pi^2\alpha^2\alpha_s^3}{5\cdot 3^8 s^6 r^4 \epsilon_{\mathrm{IR}}} 
\left[\left(720 r^8+452 r^6-696 r^4+7 r^2-15\right)S
  + \left(2880 r^{10}+368 r^8-3560 r^6+1856 r^4-56r^2\right)T\right],
\nonumber
\end{eqnarray}
\end{widetext}
where $r=2m_c/\sqrt{s}$, $\sqrt{s}$ is the center-of-mass energy,
$S=(1-4 r^2)^{1/2}$, and $T=\tanh ^{-1}S$.
It is straightforward to check that $\hat{\sigma}^{\mathrm{div}}_{1}$ and
$\hat{\sigma}^{\mathrm{div}}_{2}(J_2)$ will be canceled after including the
${\cal O}(\alpha_s)$ corrections to the S-wave LDMEs
$\langle\mathcal{O}^{\chi_{cJ}}({}^3S_{1}^{[8]})\rangle$ and
$\langle\mathcal{O}^{J/\psi}({}^3S_{1}^{[1]})\rangle$ in the SDCs of
$e^{+}e^{-}\to c\bar{c}({}^3P_{J_1}^{[8]})+c\bar{c}({}^3S_1^{[8]})$ and
$e^{+}e^{-}\to c\bar{c}({}^3S_1^{[1]})+c\bar{c}({}^3P_{J_2}^{[1]})$,
respectively.
Typical Feynman diagrams for the latter two processes are also show in
Fig.~(\ref{epem}).
Unfortunately, the standard NRQCD factorization formalism does not provide an
intrinsic mechanism to cancel $\sigma^{\mathrm{div}}_3(J_2)$, and thus it fails to
yield an IR-finite result for the cross section of
$e^{+}e^{-}\to J/\psi+\chi_{cJ}+X$ at relative order $v_c^4$.
Because $\sigma^{\mathrm{div}}_3(J_2)$ originates from the interference of
Feynman diagrams where the gluon is attached to different P-wave $Q\bar{Q}$
pairs, a feature that is independent of the initial state and does not require
the two quark pairs to have the same flavor, we conclude that there will be
similar uncanceled IR singularities in NRQCD calculations of $J/\psi$ pair and
$J/\psi$+$\Upsilon$ associated hadroproduction at NLO in $\alpha_s$.
However, the structure of the uncanceled IR singularities will generally be
more complicated there because more channels are involved.
For instance, in the case of
$gg\to c\bar{c}({}^3P_{J_c}^{[8]})+b\bar{b}({}^3P_{J_b}^{[8]})+g$, there will be
four possible interferences of the four pairings
$c\bar{c}({}^3S_1^{[1]})+b\bar{b}({}^3P_{J_b}^{[8]})$,
$c\bar{c}({}^3S_1^{[8]})+b\bar{b}({}^3P_{J_b}^{[8]})$,
$c\bar{c}({}^3P_{J_c}^{[8]})+b\bar{b}({}^3S_1^{[1]})$, and
$c\bar{c}({}^3P_{J_c}^{[8]})+b\bar{b}({}^3S_1^{[8]})$, which yield
uncanceled IR singularities.

\begin{figure}
\begin{center}
\includegraphics[width=0.48\textwidth]{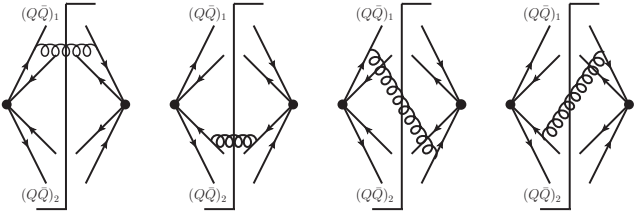}
\caption{Sample diagrams for one-loop corrections to the annihilation or
creation of two heavy-quark pairs $Q_1\bar{Q}_1$ and $Q_2\bar{Q}_2$.
The solid dots represent four-quark vertices.}
\label{NRQCD}
\end{center}
\end{figure}

The above two examples clearly demonstrate that NRQCD factorization as we know
it is spoiled by uncanceled IR singularities for inclusive production and decay
processes involving two (ore more) heavy quarkonia.
At this point, we recall that factorization implies a complete separation of
perturbative and nonperturbative effects.
In the context of the NRQCD factorization framework, one is thus led to find a
concept how to separate the problematic IR-singular terms, like
$\Gamma^{\mathrm{div}}_3(J_c)$ and $\sigma^{\mathrm{div}}_3(J_2)$, into
contributions pertaining to the hard- and soft-scale regimes.
The creation and annihilation of heavy-quark pairs clearly take place at
short distances.
To describe such processes involving two heavy-quark pairs, it is natural to
consider products of four heavy-quark fields.
Since the two heavy-quark pairs cannot be at rest simultaneously, we adopt the
covariant form of the NRQCD Lagragian, which at LO reads \cite{Manohar:1997qy}:
\begin{eqnarray}
\mathcal{L}^{\text{LO}}_{\text{NRQCD}}&=&
\bar{\psi}_{v} \left[iv\cdot D +\frac{(iD^{\mu}_{\top})(iD_{\top\mu})}{2m}
\right]\psi_v\nonumber
\\
&&{}+\bar{\chi}_v \left[-iv\cdot D +\frac{(iD^{\mu}_{\top})(iD_{\top\mu})}{2m}
\right]\chi_v,
\label{lagrangian}
\end{eqnarray}
where $m$ is the heavy-quark mass, $v^\mu=P^\mu/(2m)$ with $P^{\mu}$ being the
four-momentum of the $Q\bar{Q}$ pair, $\psi_v$ and $\chi_v$ are the
nonrelativistic four-component heavy-quark and -antiquark fields, satisfying
$\slashed{v}\psi_{v}=\psi_{v}$ and $\slashed{v}\chi_{v}=-\chi_{v}$, $D^\mu$ is
the covariant derivative, and $a_{\top}^{\mu}= a^{\mu}-v^{\mu} v\cdot a$ for
any four-vector $a^\mu$.
The relative momentum $q^{\mu}$ between $Q$ and $\bar{Q}$ corresponds to
$i\partial^{\mu}_{\top}\psi_v$.

Representative diagrams for one-loop corrections to the annihilation or
creation of two heavy-quark pairs, $Q_1\bar{Q}_1$ and $Q_2\bar{Q}_2$, are
depicted in Fig.~\ref{NRQCD}.
In the first two panels, the soft gluon interconnects the same $Q\bar{Q}$ pair.
This corresponds to the product of two four-quark operators, which separately
receive QCD corrections.
The missing link needed to remove the left-over IR singularities is depicted
in the last two panels.
Here, the two $Q\bar{Q}$ pairs cross talk by exchanging a soft gluon and so
form a joint structure, namely an eight-quark operator.

In the following, we refer to
$b\bar{b}({}^3P_{J_b}^{[8]})\to c\bar{c}({}^3P_{J_c}^{[1]})+gg$ as process A and to
$e^{+}e^{-}\to c\bar{c}({}^3P_{J_1}^{[8]})+c\bar{c}(^3P_{J_2}^{[1]})+g$ as
process B and generically denote the total and relative four-momenta of
$Q_i\bar{Q}_i$ as $P_i$ and $q_i$, respectively.
To generate the appropriate interference parts at one loop for two P-wave
$Q\bar{Q}$ states,  we need the new operators
$\psi_{b,v_1}\mathcal{K}^{\mu_1\nu_1}T^{a_1}\bar{\chi}_{b,v_1}
\bar{\psi}_{c,v_2}\gamma_{\top}^{\nu_2}T^{a_2}\chi_{c,v_2}$,
$\psi_{b,v_1}\gamma_{\top}^{\nu_1}T^{a}\bar{\chi}_{b,v_1}\bar{\psi}_{c,v_2}
\mathcal{K}^{\mu_2\nu_2}_{J_c}\chi_{c,v_2}$ and their charge conjugates for
process A and 
$\bar{\psi}_{c,v_1}\mathcal{K}^{\mu_1\nu_1}T^{a_1}\chi_{c,v_1}
\bar{\psi}_{c,v_2}\gamma_{\top}^{\nu_2}T^{a_2}\chi_{c,v_2}$,
$\bar{\psi}_{c,v_1}\gamma_{\top}^{\nu_1}\chi_{c,v_1}\bar{\psi}_{c,v_2}
\mathcal{K}^{\mu_2\nu_2}_{J_2}\chi_{c,v_2}$ and their charge conjugatates for
process B, where
$\mathcal{K}^{\mu\nu}_{0}=\frac{g^{\mu\nu}-v^{\mu}v^{\nu}}{\sqrt{3}}
(-\frac{i}{2}\overleftrightarrow{\slashed{D}}_{\top})$,
$\mathcal{K}^{\mu\nu}_{1}=\frac{-i}{2}(\overleftrightarrow
{\slashed{D}}_{\top}^{[\mu}\gamma_{\top}^{\nu]})$,
$\mathcal{K}^{\mu\nu}_{2}=\frac{-i}{2}(\overleftrightarrow
{\slashed{D}}_{\top}^{(\mu}\gamma_{\top}^{\nu)})$,
and $\mathcal{K}^{\mu\nu}=\frac{-i}{2}(\overleftrightarrow
{\slashed{D}}_{\top}^{\mu}\gamma_{\top}^{\nu})$ with
$a^{[\mu}b^{\nu]}=\frac{1}{2}(a^{\mu}b^{\nu}-a^{\nu}b^{\mu})$ and
$a^{(\mu}b^{\nu)}=\frac{1}{2}(a^{\mu}b^{\nu}+a^{\nu}b^{\mu})-\frac{g^{\mu\nu}
-v^{\mu}v^{\nu}}{3}a\cdot b$.

Using the Feynman rules derived from Eq.~(\ref{lagrangian}) in connection with
the new operators introduced above, we are now in a position to evaluate the
last two Feynman diagrams in Fig.~\ref{NRQCD}.
Although $m_c$ is about three times smaller than $m_b$, we assume that
$m_c\gg m_b v_b$ to ensure that the soft region of bottomonium is sufficiently
separated from the hard region of charmonium so that the nonrelativistic
approximation still applies to the latter.
Working in covariant gauge, we show that the results are gauge independent.
The details of our calculation will be presented elsewhere \cite{hkw}.
For space limitation, we here merely explain how to perform the loop
integrations, taking the third diagram in Fig.~\ref{NRQCD} as an example and
working in Feynman gauge.
The arising loop integral reads 
\begin{eqnarray}
  I &=& -ig_s^2\mu^{4-D}\int \frac{d^D l}{(2\pi)^D}
  \nonumber\\
&&\times
  \frac{
    \left[v_1+ \frac{(2q_{1}+l)_{\top}}{2m_1}\right]\cdot
   \left[v_2+ \frac{(2q_{2}+l)_{\top}}{2m_2}\right]}
{l^2\left[l\cdot v_1+\frac{(l+q_1)_{\top}^{2}}{2m_1}\right]
  \left[l\cdot v_2+\frac{(l+q_2)_{\top}^{2}}{2m_2}\right]}.
\end{eqnarray}
Expanding the heavy-quark propagators in $1/m_i$ and dropping terms of
order $1/m_i^2$ and higher, which contribute at higher orders in
$v_i^2$, we obtain
\begin{eqnarray}
I&=&-\frac{i g_s^2 \mu^{4-D}}{m_1m_2}\int \frac{d^D l}{(2\pi )^D}\frac{1}{l^2}
\left[\frac{q_{1}\cdot {q_2}}{(l\cdot v_1) (l\cdot v_2)}
\right.
\nonumber\\
&&{}-\frac{(l\cdot q_1)(v_1\cdot q_2)}{(l\cdot v_1)^2 (l\cdot v_2) }
-\frac{ (l\cdot q_2)(v_2\cdot q_1)}{(l\cdot v_1)(l\cdot v_2)^2 }
\nonumber\\
&&{}+\left.\frac{ (v_1\cdot v_2) (l\cdot q_1)(l\cdot q_2)}
{(l\cdot v_1)^2 (l\cdot v_2)^2}\right]+I_0
\nonumber\\
&=& \frac{\alpha_s \mu^{4-D}}{\pi m_1m_2}
\left(\frac{1}{\epsilon_{\rm UV}}-\frac{1}{\epsilon_{\rm IR}}\right)
\left[q_1\cdot q_2
\vphantom{
\frac{(\omega^2+2)\sqrt{\omega^2-1}-3\omega\ln(\sqrt{\omega^2-1}+\omega)}{ 2(\omega^2-1)^{5/2}}
}
\right.
\label{eq:i}
\\
&&{}\times
\frac{\ln(\omega+\sqrt{\omega^2-1})-\omega\sqrt{\omega^2-1}}{ 2(\omega^2-1)^{3/2}}
+(v_1\cdot q_2) (v_2\cdot q_1)
\nonumber\\
&&{}\times\left.
\frac{(\omega^2+2)\sqrt{\omega^2-1}-3\omega\ln(\sqrt{\omega^2-1}+\omega)}{ 2(\omega^2-1)^{5/2}}\right]+I_0,
\nonumber
\end{eqnarray}
where $\omega=v_1\cdot v_2$ and $I_0$ includes irrelevant terms that will cancel
in the sum over all diagrams.
The ultraviolet singularities are removed via operator renormalization.
Multiplying Eq.~(\ref{eq:i}) with the corresponding SDCs, which is the
interference of $b\bar{b}({}^3P_{J_b}^{[8]})\to c\bar{c}({}^3S_{1}^{[8]})+g$ 
and $b\bar{b}({}^3S_1^{[8]})\to c\bar{c}({}^3P_{J_c}^{[1]})+g$ for process A and
the interference of 
$e^{+}e^{-}\to c\bar{c}({}^3P_{J_1}^{[8]})+c\bar{c}({}^3S_{1}^{[8]})$ and
$e^{+}e^{-}\to c\bar{c}({}^3S_1^{[1]})+c\bar{c}({}^3P_{J_2}^{[1]})$ for process B,
and decomposing the tensor and color structure into the basis of the
total-angular-momentum and color states, we find that the IR-singular parts
exactly match those in
$\Gamma^{\mathrm{div}}_3(J_c)$ and $\sigma^{\mathrm{div}}_3(J_2)$.
We wish to emphasize that the loop integrals are process independent although
they depend on $v_1\cdot v_2$.

In summary, we discovered a surprising loophole in the standard formulation of
the NRQCD factorization approach which manifests itself in the failure of IR
cancelation in the presence of two (or more) P-wave $Q\bar{Q}$ Fock states.
This inevitably causes NRQCD factorization to break down for any decay or
production process involving two (or more) heavy quarkonia at a certain order
of $v_Q^2$.
We illustrated this for two phenomenologically important example processes,
$\Upsilon \to \chi_{cJ}+X$ and $e^+e^-\to J/\psi+\chi_{cJ}+X$, at NLO in
$\alpha_s$.
As a solution to this problem, we introduced new types of operators and
demonstrated that their NLO corrections precisely reproduce the uncanceled IR
singularities, which may thus be attributed to the soft regime of NRQCD.
This implies that it is possible to generalize the factorization formalism
within the very same NRQCD effective field theory so as to allow for the
successful theoretical description of processes involving two (or more) heavy
quarkonia.
The explicit construction of such a generalized NRQCD factorization formalism
and its applications to heavy-quarkonium phenomenology are left for future
work.

We thank G.~T.~Bodwin and E.~Braaten for very useful comments.
This work was supported in part by
BMBF
Grant No.\ 05H15GUCC1.
The work of X.P.W. was supported in part by CSC Scolarship No.\ 201404910576.




\end{document}